\documentclass[a4paper,showpacs,twocolumn,amsmath,amssymb,floatfix,prb,preprintnumbers,footinbib]{revtex4}
\usepackage[pdftex]{graphicx}%, superscriptaddress
\usepackage{dcolumn}% Align table columns on decimal point
\usepackage{bm}
\usepackage{dsfont}
\usepackage{color}
\usepackage{url}
\usepackage[colorlinks=true,citecolor=blue]{hyperref}
\usepackage{amsmath,amssymb}
\newcommand{\ttt}{$\mathcal{T}_3$~}
\newcommand{\ttm}{\mathcal{T}_3}
\newcommand{\ii}{\textrm{i}}

\begin{document}

\title{Barrier transmission of Dirac-like pseudospin-one particles}

\author{Daniel F. Urban}
\email{urban@physik.uni-freiburg.de}
\affiliation{Physikalisches Institut, Albert-Ludwigs-Universit\"at, D-79104 Freiburg, Germany}
\author{Dario Bercioux}
\affiliation{Physikalisches Institut, Albert-Ludwigs-Universit\"at, D-79104 Freiburg, Germany}
\affiliation{Freiburg Institute for Advanced Studies, Albert-Ludwigs-Universit\"at, D-79104 Freiburg, Germany}
\author{Michael Wimmer}
\affiliation{Instituut-Lorentz, Universiteit Leiden, P.O. Box 9506, 2300 RA Leiden, The Netherlands}
\author{Wolfgang H\"ausler}
\affiliation{Institut f\"ur Physik, Universit\"at Augsburg, D-86135 Augsburg, Germany}

\begin{abstract}

We address the problem of barrier tunneling in the two-dimensional \ttt lattice (dice lattice).
In particular we focus on the low-energy, long-wavelength approximation for the Hamiltonian of the system,
where the lattice can be described by a Dirac-like Hamiltonian associated with a pseudospin one.
The enlarged pseudospin $S = 1$ (instead of S = $1/2$ as for graphene) leads
to an enhanced ``super" Klein tunneling through rectangular electrostatic barriers.
Our results are confirmed via numerical investigation of the tight-binding model of the lattice.
For a uniform magnetic field, we discuss the Landau levels and we investigate the transparency of a rectangular magnetic barrier.
We show that the latter can mainly be described by semiclassical
orbits bending the particle trajectories, qualitatively similar
as it is the case for graphene.
This makes it possible to confine particles with magnetic barriers of sufficient width.
\end{abstract}

\date{\today}


% 73.23.Ad   Ballistic transport
% 73.40.Gk   Tunneling
% 73.23.-b   Electronic transport in mesoscopic systems
% 73.22.Pr   Electronic structure of graphene

\maketitle

\section{Introduction}

Nanostructures in two-dimensional electron systems are usually patterned by using
gate electrodes  creating adequate electrostatic (scalar) potential barriers.
Typical examples include quantum wires and quantum dots in semiconductor heterostructures.
Yet this procedure is not applicable in the case of graphene, a single layer of carbon atoms
arranged in a honeycomb lattice (HCL).~\cite{neto:2009,Beenakker:2008} At low energies, the HCL can be described by a
relativistic Dirac-Weyl Hamiltonian.~\cite{DW:ref}  In this limit an electrostatic potential barrier of arbitrary height
and thickness is fully transparent for low-energy electrons at certain incident angles.\cite{katsnelson:2006}
This effect | known as Klein tunneling~\cite{klein} | has recently been demonstrated experimentally
using graphene heterostructures.~\cite{klein:exp}
On the other hand, sufficiently wide magnetic barriers yield zero transparency and therefore
make it possible to confine electrons.~\cite{demartino:2007,mag:confinement}

In this article we investigate confining properties of the so-called \emph{dice}- or
${\cal T}_3$-lattice~\cite{sutherland:1986,vidal:1998}, c.f. Fig.~\ref{fig:Tau3lattice}a.
This lattice is described by a Dirac-like
Hamiltonian~\cite{bercioux:2009,bercioux:2011} | cf.~\eqref{model} below | resembling the one
describing the HCL and accordingly graphene, but with an enlarged pseudospin
$\:S=1\:$~\cite{bercioux:2009} compared to $\:S=\frac{1}{2}\:$ for graphene.~\cite{neto:2009}
It turns out that the presence of an integer pseudospin $S=1$ has
striking consequences on the Klein tunneling:
(i) Electrostatic barriers become even more transparent compared to the HCL case and (ii) there is a regime
of super-Klein-tunneling with perfect transmission independent of the angle
of incidence. The latter is related
to a negative refraction index $\nu=-1$ and allows for the realization of a perfect focusing system.

The paper is organized as follows. First, in Sec.~\ref{sec:low:en}, we introduce the low energy description of the ${\cal T}_3$-lattice
and its implications on scattering at a barrier are presented. In Sec.~\ref{sec:el:bar} we solve the scattering problem of an electrostatic barrier
and discuss the effect of Klein tunneling.
In Sec.~\ref{sec:num:sol} we compare the analytical results with a numerical calculation using the recursive Green's function method.
Section~\ref{sec:LLs} is devoted to the general features
of the ${\cal T}_3$-lattice in presence of a uniform magnetic field and Sec.~\ref{Sec:MagneticBarrier} deals with the scattering problem of a
magnetic barrier.

\section{Low energy description of the ${\cal T}_3$-lattice\label{sec:low:en}}

\begin{figure}
	\begin{center}
	\includegraphics[width=0.9\columnwidth]{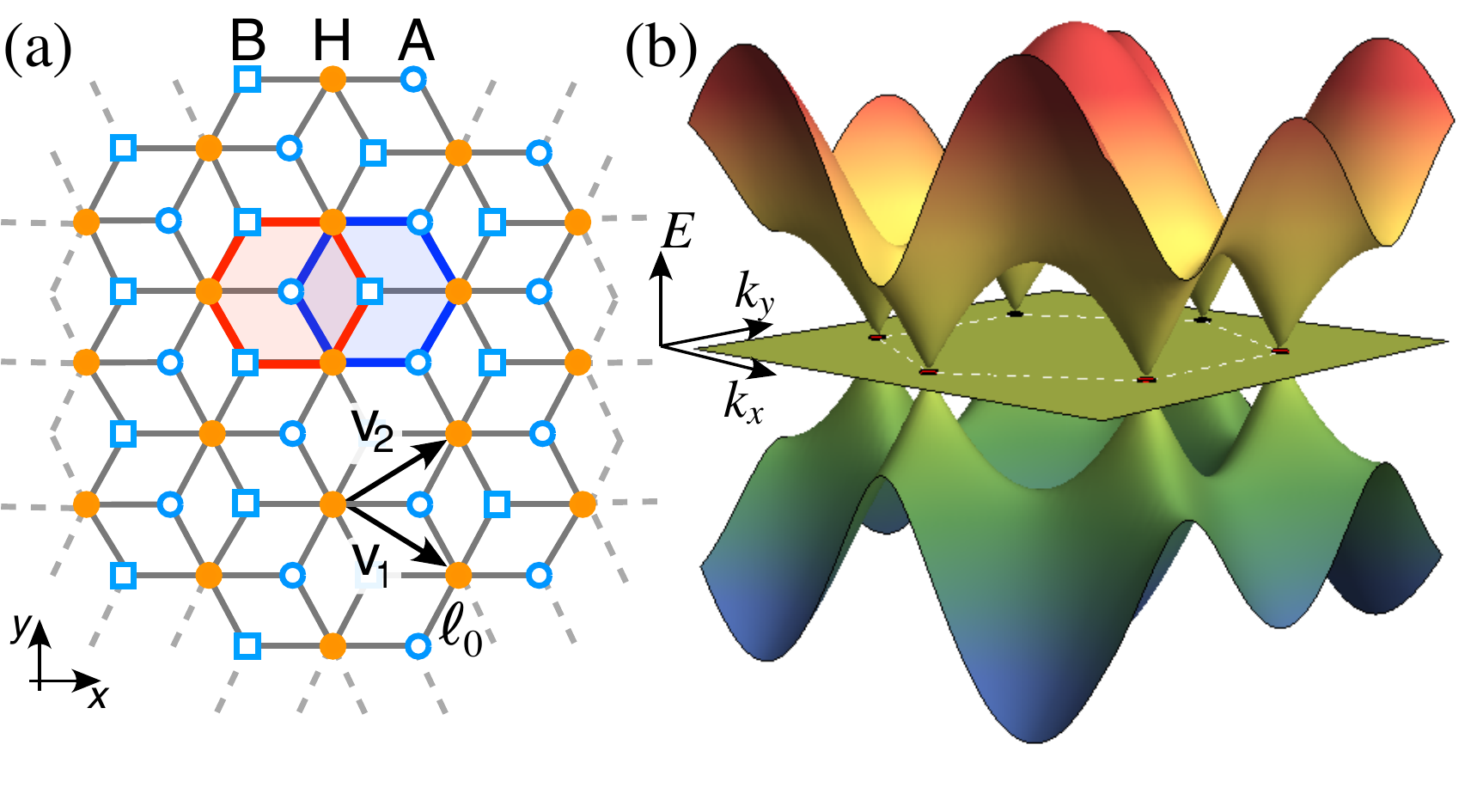}
	\caption{\label{fig:Tau3lattice} (a) The \ttt\--lattice in the $x$--$y$--plane,
    characterized by translation vectors $\mathbf{v}_1=(3/2;-\sqrt{3}/2)a_0$ and $\mathbf{v}_2=(3/2;\sqrt{3}/2)a_0$, with lattice constant $a_0$.
    Rim sites A and B (open circles and squares, respectively) have a lower connectivity of 3 compared to 6 for the hub site (solid circles).
    The \ttt lattice can be viewed as two nested HCLs (emphasized by the two shaded hexagons).
    (b) The energy spectrum of the bulk \ttt\ lattice. The first Brillouin zone is indicated by the dashed line on the flat band.}
	\end{center}
\end{figure}

The \ttt lattice [cf.\ Fig.~\ref{fig:Tau3lattice}a] contains three inequivalent sites per unit cell.
Two of these lattice sites (generally called \emph{rim} sites A and B) are
threefold coordinated while the third
site (referred to as \emph{hub}~H) is connected to six nearest neighbors. The energy spectrum of the \ttt\ lattice
consists of two electron-hole symmetric branches, in addition to a unique non-dispersive band at the charge
neutrality point.~\cite{bercioux:2009} The latter roots in the lattice topology, which allows for insulating states
with finite wave function amplitudes on the rim sites and vanishing amplitudes on the hubs.
The reciprocal lattice of \ttt\ has a hexagonal first Brillouin zone [cf.\ Fig.~\ref{fig:Tau3lattice}b]
with the electron-hole symmetric bands touching in the six corners of the hexagon (Dirac points).
As in graphene, there are two inequivalent Dirac points, labeled $K$ and $K'$.

%
%
%%%%%%%%%%%%
%%%%%%%%%%%%
%
%

The low-energy Hamiltonian for the ${\cal T}_3$ lattice
is obtained by a long-wavelength approximation to the tight-binding Hamiltonian.
For a given $K$ point it reads\cite{bercioux:2009}
%
%
%%%%%%%%%%%%%%%%%
\begin{equation}\label{model}
\mathcal{H}=v_\text{F}\bm{S}\cdot\left(-\ii\bm{\nabla}-\frac{q}{c}\bm{A}\right)+V
\end{equation}
%%%%%%%%%%%%%%%%%
%
%
(in units where $\hbar\equiv1$).
Here $v_{\rm F}$ is the Fermi velocity,  the operator $\bm{\nabla}=(\partial_x,\partial_y)$ is the
differential operator in the $xy$-plane, and we consider particles with negative charge $q=-e<0$.
Moreover, $\:V(x,y)\:$ represents a scalar electrostatic potential (proportional to a unit matrix in spinor
space), while $\:\bm{A}(x,y)\:$ is the vector potential related to the
$z$-component of a magnetic field.
Pseudospin components~\cite{bercioux:2009} of $\bm{S}$,
%
%
%%%%%%%%%%%
\begin{align}\label{eq:gm}
 \footnotesize S_x & =\frac{1}{\sqrt{2}}
 \begin{pmatrix}
 0 & 1 & 0 \\
 1 & 0 & 1 \\
 0 & 1 & 0
 \end{pmatrix},\
 S_y=\frac{1}{\sqrt{2}} \begin{pmatrix}
 0 & -\ii & 0 \\
 \ii & 0 & -\ii \\
 0 & \ii & 0
 \end{pmatrix},\ \\
\nonumber  & ~~~~~~~~~ S_z =\begin{pmatrix}
 1 & 0 & 0 \\
 0 & 0 & 0 \\
 0 & 0 & -1
 \end{pmatrix},
\end{align}
%%%%%%%%%%%%%%%%%%%%
%
%
form a three-dimensional representation of SU(2) and
satisfy angular momentum commutation relations $[S_m,S_n]=\ii
\epsilon_{mn\ell}S_\ell$.
However, contrary to Pauli matrices,
they do not form a Clifford algebra, \emph{i.e.},\ $\{S_n,S_m\}\neq 2 \delta_{n,m} \mathbb{I}_3$.

The ${\cal T}_3$-model (\ref{model})
generalizes graphene by enlarging pseudospin. Each plane-wave
momentum eigenstate $\:|\bm{p}\rangle\:$ (at $V=0$ and
$\bm{A}=\bm{0}$) is pseudospin {\em polarized} with respect to the
$\bm{p}$-direction along which it experiences three possible
eigenvalues $\:S_{\bm{p}}=(+1,0,-1)\:$. They translate into
pseudo-Zeeman split energies $\:E=v_{\rm F}(+p,0,-p)\:$ constituting three
bands. The ``non-magnetic'' level $S_{\bm{p}}=0$ comprises the
topologically protected dispersionless band.

In the following, we investigate the transmission probability through electrostatic or magnetic
rectangular barriers of constant height and thickness $d$ which
we assume as homogeneous along the $y$-direction as the
archetype of a tunneling obstacle (c.f. Fig.~\ref{fig:setup}).
Accordingly, the three-component eigenspinors of \eqref{model} have the structure
%
%
%%%%%%%%%%
\begin{align}\label{general:wf}
\:\psi(x,y)=(\psi_{\rm A}(x),\psi_{\rm H}(x),\psi_{\rm
B}(x)){\rm e}^{{\rm i}k_yy}\:
\end{align}
%%%%%%%%%%%
%
%
which comprises the amplitudes on the three sublattices of the ${\cal
T}_3$ lattice, $\psi_{\rm A},\psi_{\rm H}$, and $\psi_{\rm B}$.

In order to obtain the transmission probability at non-zero $V$
or $\bm{A}$, we need to solve for the scattering problem by
matching wave functions across the rectangular barrier, c.f.\
Fig.~\ref{fig:setup}. Strictly sharp potential steps would
cause large momentum scattering and so violate our assumption
of close vicinity to one chosen $K$-point. Therefore, as is
customary,\cite{katsnelson:2006} we assume ultimately potential
variations as smooth on the length scale of the lattice constant
$a_0$ but as sharp on the Fermi wavelength
$\lambda_\text{F}=2\pi v_{\rm F}/|E|$ which is large at low
energies $|E|$.

To this end, let us derive the adequate matching conditions
for $\:S=1\:$ particles obeying \eqref{model} that turn out to be
different from Schr\"odinger particles and different from
$\:S=\frac{1}{2}\:$ Dirac particles. In the latter case
continuity of the wave function is required across a potential
step (together with continuity of the spatial derivative in the
Schr\"odinger case). These familiar conditions need to be
altered when dealing with the Hamiltonian \eqref{model}. As
usual, we integrate the eigenvalue equation $\:H\psi=E\psi\:$
over the small interval $\:x\in[-\epsilon,\epsilon]\:$ along the
$x$-direction and let $\epsilon$ eventually go to zero. This
yields
%
%
%%%%%%%%%%%%%%%%
\begin{subequations}\label{matching}
\begin{align}
\psi_{\rm H}(\epsilon)& = \psi_{\rm H}(-\epsilon)\\
\psi_{\rm A}(\epsilon)+\psi_{\rm B}(\epsilon) & =
\psi_{\rm A}(-\epsilon)+\psi_{\rm B}(-\epsilon)
\end{align}
\end{subequations}
%%%%%%%%%%%%%%%%
%
%
for non-diverging scalar or vector potentials, $V$ and $\bm{A}$.
While $\psi_{\rm H}$ must be continuous, Eq.~\eqref{matching}
only demands continuity for $\psi_{\rm A}+\psi_{\rm B}$, allowing
still for redistribution of occupation
probability between A and B rim sites across potential steps.
A similar integration of $\:H\psi=E\psi\:$ along the $y$-direction yields
continuity of $\psi_{\rm H}$ and of $\psi_{\rm A}-\psi_{\rm B}$
as matching conditions along the $x$-direction.

A discontinuity in the wave function components $\psi_{\rm A/B}$ seems unexpected
and an explanation in terms of physical quantities is helpful. Therefore, we calculate
the probability current
%
%
%%%%%%%%%%%%%%%%%
\begin{eqnarray}
    \bm{j}&=&\left(
    \begin{array}{c}
        j_x\\
        j_y
    \end{array}
    \right)
    \;=\;\sqrt{2}v_{\text F}\left(
    \begin{array}{c}
        \rm{Re}\left[\psi_{\rm H}^*\left(\psi_{\rm A}+\psi_{\rm B}\right)\right]\\
        \rm{Im}\left[\psi_{\rm H}^*\left(\psi_{\rm A}-\psi_{\rm B}\right)\right]
    \end{array}
    \right)
\end{eqnarray}
%%%%%%%%%%%%%%%%%%
%
%
(see Appendix~\ref{App:one}), that satisfies a continuity equation,
%
%
%%%%%%%%%%%%%%%
\begin{eqnarray}
    \frac{\partial}{\partial t}|\psi(x,t)|^2+\nabla\cdot\bm{j}=0,
\end{eqnarray}
%%%%%%%%%%%%%%
%
%
for the probability density $|\psi|^2=|\psi_{\rm A}|^2+|\psi_{\rm B}|^2+|\psi_{\rm H}|^2$.
From the expressions of $j_x$ and $j_y$ we can conclude, that the matching conditions
correspond to a conservation of the probability current perpendicular to the barrier.
On the other hand, parallel to the barrier the currents need not be
equal on either side.

More generally, for a barrier perpendicular to the vector ${\bm n}=(\cos(\alpha),\sin(\alpha))^\text{T}$, with $\alpha\in(-\frac{\pi}{2},\frac{\pi}{2})$, the current
%
%
%%%%%%%%%%%%%%%
\begin{equation}
    {\bm j}\cdot{\bm n}=\sqrt{2}v_{\text F}\rm{Re}\left[\psi_{\rm H}^*\left(\psi_{\rm A}\rm{e}^{-i\alpha}+\psi_{\rm B}\rm{e}^{i\alpha}\right)\right]
\end{equation}
%%%%%%%%%%%%%%%%%%%
%
%
is conserved across the barrier. Therefore, only the linear combination
$(\psi_{\rm A}\rm{e}^{-i\alpha}+\psi_{\rm B}\rm{e}^{i\alpha})$ is continuous across the barrier while
$(\psi_{\rm A}\rm{e}^{-i\alpha}-\psi_{\rm B}\rm{e}^{i\alpha})$, in general, is discontinuous.

\section{Electrostatic barrier\label{sec:el:bar}}
%
%
%%%%%%%%%%%%
\begin{figure}
	\begin{center}
	\includegraphics[width=\columnwidth]{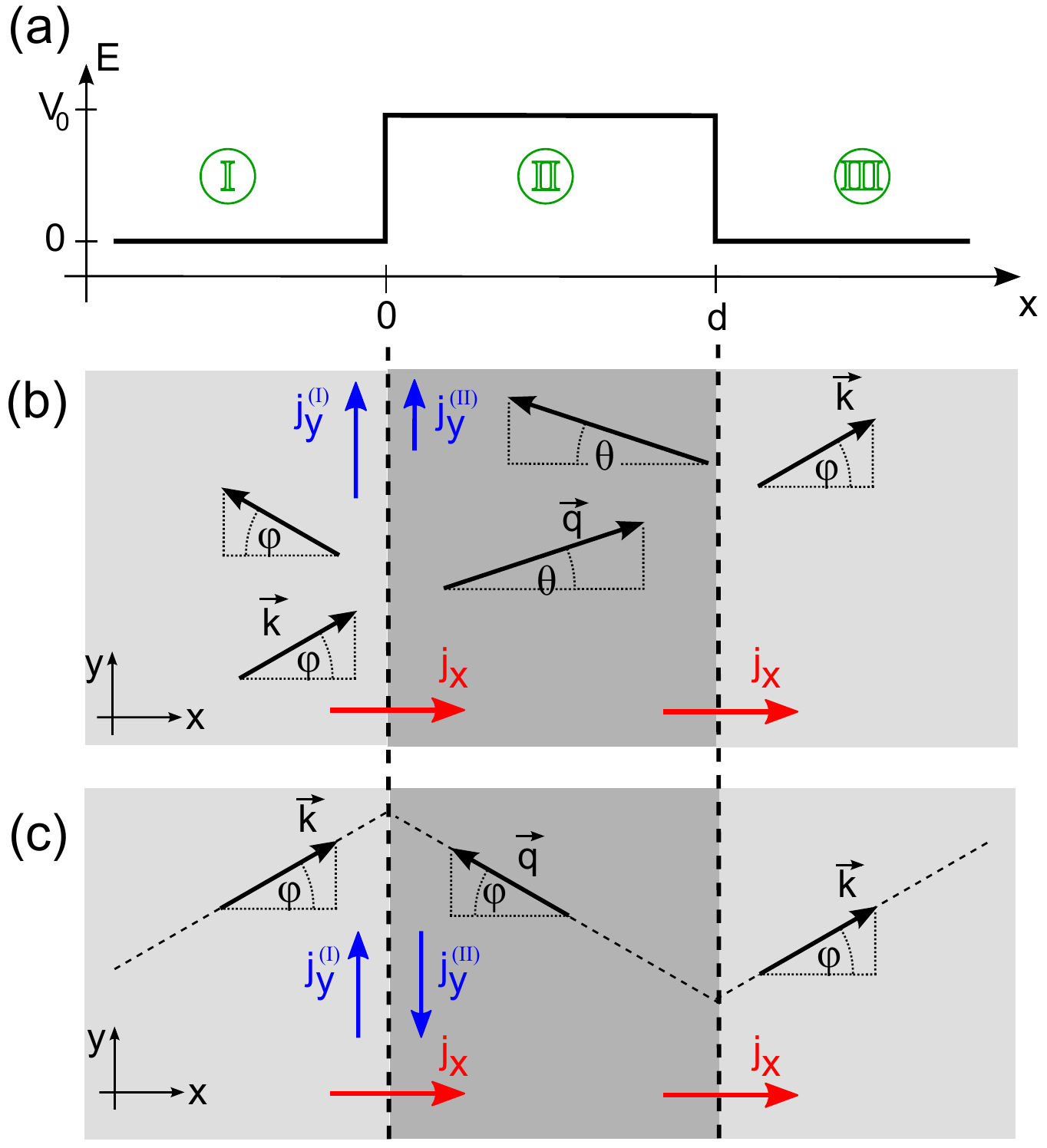}
	\caption{\label{fig:setup} (a) Sketch of the scattering region with an electrostatic barrier of width $d$.
    (b) Wave vectors corresponding to the different plane-wave components in $\psi_\text{I}$, $\psi_\text{II}$, and $\psi_\text{III}$; see Eqs.\ (\ref{psi1})--(\ref{psi3}), for $E>V_0$.
    (c) Scattering trajectory in the special case $E=V_0/2$. Note that the opposite orientation of $\bf q$ in the barrier region is due to propagation of holes [$(E-V_0)<0$] in II, compared to electrons ($E>0$) in I and III. }
    \end{center}
\end{figure}
%%%%%%%%%%%%
%
%

Let us consider the case of a rectangular electrostatic barrier
%
%
%%%%%%%%%%%%%%%%%%
\begin{equation}\label{vbarrier}
V(x,y)=V_0\Theta(x)\Theta(d-x)
\end{equation}
%%%%%%%%%%%%%%%%%%
%
%
of width $d$ and height $V_0$ (cf.\ Fig.\ \ref{fig:setup}). Here $\Theta(x)$ is the Heaviside
step function. Due to momentum conservation parallel to
the barrier, the wave function can be written as
$\:\psi(x)\text{e}^{{\rm i}k_yy}\:$ [cf.\ Eq.\ \eqref{general:wf}], with homogenous $y$-component $k_y$
of the wave vector. For $\psi(x)$ we assume in- and out-going plane waves at $x<0$,
%
%
%%%%%%%%%%%%%%%%
\begin{equation}
\label{psi1}
    \psi_{\rm I}(x) =
    \left(\begin{array}{c}
        \text{e}^{-{\rm i}\varphi}\\
        \sqrt{2}s\\
        \text{e}^{{\rm i}\varphi}
    \end{array}\right)
    \text{e}^{{\rm i}k_x x}
    +r\left(\begin{array}{c}
        \text{e}^{{\rm i}\varphi}\\
        -\sqrt{2}s\\
        \text{e}^{-{\rm i}\varphi}
    \end{array}\right)
    \text{e}^{-{\rm i}k_x x}
\end{equation}
%%%%%%%%%%%%%%%%
%
%
with reflection amplitude $r$,
that are solutions to \eqref{model} at energy $E\neq0$.
Consequently, the wave function under the barrier,
$\:0\le x\le d\:$,
at non-zero $V$ is of the form
%
%
%%%%%%%%%%%%%%%%%
\begin{equation}
\label{psi2}
    \psi_{\rm II}(x)=
    a\left(\begin{array}{c}
        \text{e}^{-{\rm i}\theta}\\
        \sqrt{2}s'\\
        \text{e}^{{\rm i}\theta}
    \end{array}\right)
    \text{e}^{{\rm i}q_x x}
    +b\left(\begin{array}{c}
        \text{e}^{{\rm i}\theta}\\
        -\sqrt{2}s'\\
        \text{e}^{-{\rm i}\theta}
    \end{array}\right)
    \text{e}^{-{\rm i}q_x x}\;,
\end{equation}
%%%%%%%%%%%%%%%%
%
%
while at $x>d$ we consider out-going plane waves,
%
%
%%%%%%%%%%%%%%%%%%
\begin{equation}
\label{psi3}
    \psi_{\rm III}(x)=
    t\left(\begin{array}{c}
        \text{e}^{-{\rm i}\varphi}\\
        \sqrt{2}s\\
        \text{e}^{{\rm i}\varphi}
    \end{array}\right)
    \text{e}^{{\rm i}k_xx}
\end{equation}
%%%%%%%%%%%%%%%%%%
%
%
of transmission amplitude $t$. In (\ref{psi1})--(\ref{psi3})
$\:\varphi=\arctan(k_y/k_x)\:$ is the direction of incoming and
outgoing wave vectors. The barrier alters the wave vector ${\bm
q}=(q_x,k_y)$ in its $x$-component,
$\:q_x=\sqrt{(E-V_0)^2-k_y^2}\:$, and direction
$\:\theta=\arctan(k_y/q_x)\:$. Further, we define
$\:s=\text{sign}(E)\:$ and $\:s'=\text{sign}(E-V_0)$, and assume
$E\neq V_0$; the special case $E=V_0$ is treated separately
below.

By imposing the matching conditions \eqref{matching} to $\psi_{\rm I}$
and $\psi_{\rm II}$ at $x=0$ and to $\psi_{\rm II}$ and
$\psi_{\rm III}$ at $x=d$ we obtain a set of linear equations for the parameters
$r,t,a,b$. Solving these equations yields
%
%
%%%%%%%%%%%%%%%%%
\begin{align}
\label{vsolutions}
    r&=-\left(1-\gamma^2\right)\left(1-\text{e}^{2{\rm i}qd}\right){\cal D}^{-1}
        \nonumber\\
    a&= -2\gamma\left(1+ss'\gamma\right){\cal D}^{-1}
        \nonumber\\
    b&= 2\gamma\left(1-ss'\gamma\right)\text{e}^{2 {\rm i}qd}{\cal D}^{-1}
        \nonumber\\
    t&= -4 ss'\gamma\,\text{e}^{-{\rm i}(k-q)d}{\cal D}^{-1}
\end{align}
%%%%%%%%%%%%%%%%
%
%
with $\:\gamma=\cos\varphi/\cos\theta\:$ and $\:{\cal
D}=\text{e}^{2{\rm i}qd}\left(1-ss'\gamma\right)^2-
\left(1+ss'\gamma\right)^2\:$. Finally, from Eqs.~\eqref{vsolutions}
we obtain the transmission probability
%
%
%%%%%%%%%%%%%%%%
\begin{equation}
\label{vtransmission}
    T_{\ttm}=|t|^2=\frac{\gamma^2}{\gamma^2+\frac{1}{4}\left(1-\gamma^2\right)^2\sin^2(qd)}\,.
\end{equation}
%%%%%%%%%%%%%%%%%
%
%
This result coincides with the corresponding expression obtained recently for the line-centered square (Lieb) lattice.~\cite{Shen:2010}
The functional form~\eqref{vtransmission} differs considerably from the graphene case for the same rectangular
barrier,~\cite{neto:2009}
%
%%%%%%%%%%%%%%%
\begin{equation}
\label{eq:T:HCL}
    T_\text{HCL}=\left[1 + \sin^2(q d) \left(\frac{s' \sin(\theta) - s \sin(\varphi)}
    {\cos(\theta) \cos(\varphi)}\right)^2\right]^{-1}\;.
\end{equation}
%%%%%%%%%%%%%%%

\begin{figure}
\begin{center}
    \includegraphics[width=\columnwidth]{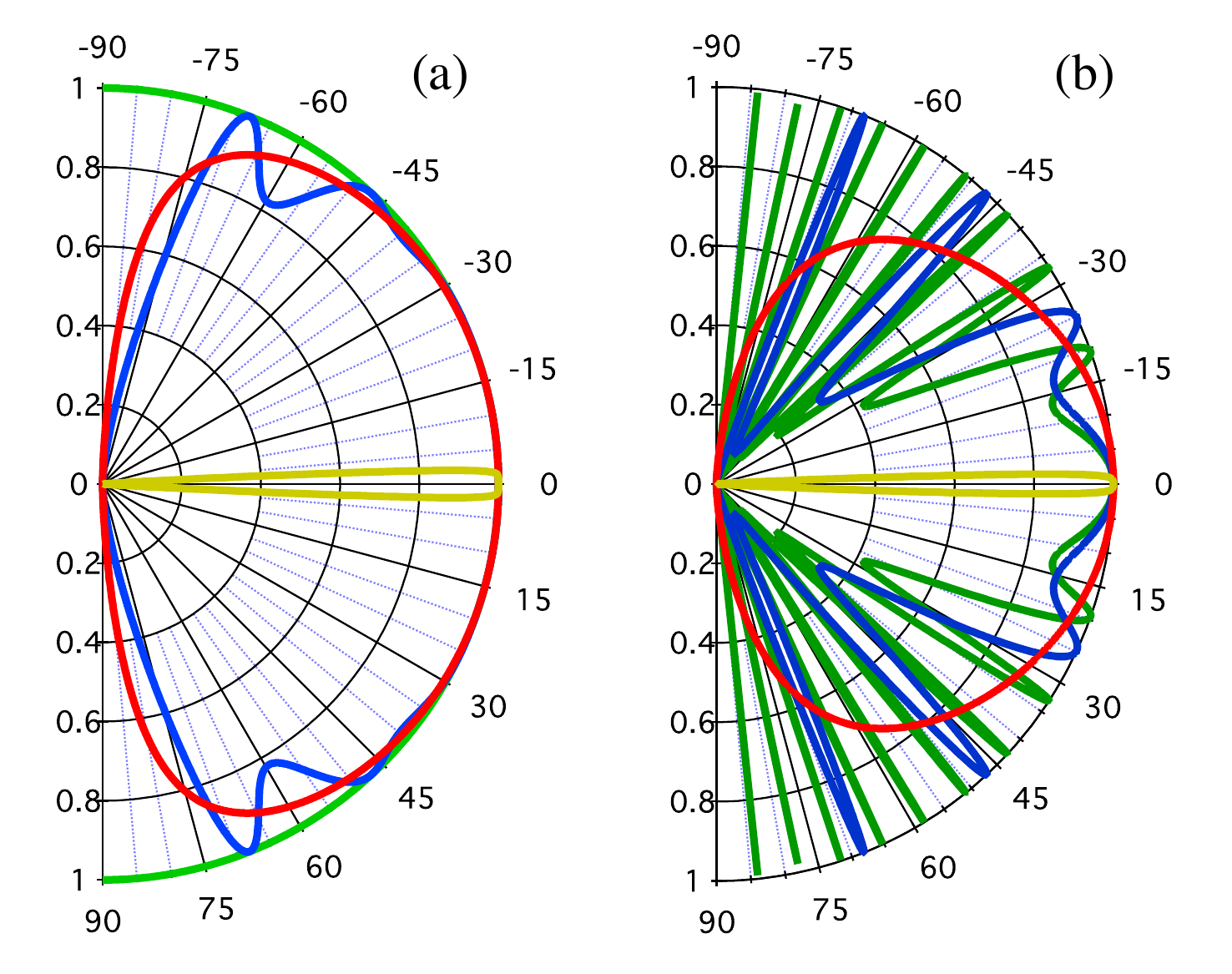}
    \caption{Polar plot of the transmission probability $T$ as a function of the injection angle $\varphi$ for
    different values of the energy $E$ for (a) the \ttt\ lattice  and for (b) the HCL lattice.
    The width of the barrier is $d V_0=60$ in both panels. The red dashed line is for $E=0.1V_0$,
    the blue dotted line is for $E=0.4V_0$, the green solid line is for $E=0.5V_0$, and the black dot-dashed line is for $E=0.95V_0$.
    For the \ttt\ lattice at energy $E=0.5V_0$ we get perfect transmission, independent of the
    incident angle $\varphi$.\label{fig:transmission}}
\end{center}
\end{figure}

The transmission $T_{\ttm}$ is plotted in Fig.~\ref{fig:transmission}(a) versus the incident angle $\varphi$ for various energies
$\:0\le E<V_0\:$, \emph{i.e.}, in the regime of $ss'=-1$. Figure~\ref{fig:transmission}(b) shows the result (\ref{eq:T:HCL})
for the HCL for the same parameters for comparison.
For the $\ttm$-case, we find almost perfect transmission for a wide range of incident angles.
For perpendicular incidence, \emph{i.e.}, $\varphi=0$ which implies $\theta=0$, we always observe perfect transmission $T=1$ that is usually referred to as `Klein tunneling'. The condition $T=1$ is also obtained for resonant values of $qd$
being equal to an integer multiple of $\pi$, as in graphene. Comparison between the two cases shows that
electrostatic barriers on ${\cal T}_3$ are significantly more transparent than on graphene.
We can even establish the inequality
%
%
%%%%%%%%%%%%%%
\begin{equation*}
T_{\ttm}\ge T_{\tiny\rm HCL}\quad\mbox{for}\quad E\in\left(0,\frac{V_0}{2}\right),
\end{equation*}
%%%%%%%%%%%%%
%
%
holding at all $\varphi$ for same barrier height $V_0$ and
thickness $d$. We conjecture that
these high transparencies make ${\cal T}_3$ even ``more ballistic'' than graphene and less sensitive to disorder~\cite{disorderGraphene}
in its transport properties at zero magnetic field.
%
%
%%%%%%%%%%%%%%
%%%%%%%%%%%%%%
%
%

The most striking feature in the transmission probability is that at energy $E=V_0/2$, where $\gamma=1$, we obtain
perfect transmission $T=1$ for {\em any} angle $\varphi$ of incidence and {\em any} barrier thickness.
This physical effect can be analyzed by considering the scattering of a plane wave at a single electrostatic interface $V(x,y)=V_0\Theta(x)$
which describes a \emph{np}-junction. At $E=V_0/2$ electrons
propagating at $x<0$ and incoming angle $\varphi$ continue to propagate
as holes at $x>0$ and a diffraction angle $\theta=\pi-\varphi$ [cf.~Fig.\ \ref{fig:setup}(c)]. Therefore, the system can be characterized by a negative diffraction index $n$:\cite{Cheianov:2007}
%
%
%%%%%%%%%%%
\begin{equation}
\label{negative:n}
    n=s\,s'\frac{\sin\varphi}{\sin\theta}=-1\,.
\end{equation}
%%%%%%%%%%%
%
%
This holds true for both, the \ttt and the HCL lattice.
However, for the HCL the reflection probability keeps an angular dependence
%
%
%%%%%%%%%%
\begin{equation*}
    R_\text{HCL}=\frac{1}{2} [1 - \cos(2 \varphi)]\,
\end{equation*}
%%%%%%%%%%%%%%%%%
%
%
while in the case of the \ttt\-lattice $R_\text{\ttt}=0$.

Finally, we discuss the special case of \mbox{$E=V_0$}. Here
let us first consider $k_y=0$, that is, $\varphi=0$.
Then the wave function in the barrier region is constant,
\begin{equation}
    \Psi_\text{II}(x)=(a,h,b)^T.
\end{equation}
The requirement of current conservation yields the four equations
%
%
%%%%%%%%%%%%%%%%%%
\begin{eqnarray}
    2(1+r)=a+b,&\qquad& a+b=2t\text{e}^{\text{i} k_x d}, \\
    \sqrt{2}s(1-r)=h, &\qquad& h =\sqrt{2}\ s\, t\,\text{e}^{\text{i} k_x d}
\end{eqnarray}
%%%%%%%%%%%%%%%%%
%
%
which readily give $r=0$, $(a+b)=2$, $h=\sqrt{2}s$, and $t=\text{e}^{-\text{i}k_x d}$ and therefore perfect transmission, $|t|^2=1$.
Note that $(a-b)$, describing a finite current in the $y$-direction in the barrier region,
\emph{a priori} cannot be determined.
We may even include an arbitrary linear combination of topological states,
%
%
%%%%%%%%%%%
\begin{eqnarray}
\label{eq:topStates}
    \Psi_\text{II}^\text{top}(x)&=&\sum_{q}\alpha_{q}
    \begin{pmatrix}
        1\\
        0\\
        -1
    \end{pmatrix}
    \text{e}^{{\rm i}q x}
\end{eqnarray}
%%%%%%%%%%%%%
%
%
which automatically satisfies $\psi_{\text A}+\psi_{\text B}=0$
for any $q$ and any amplitudes $\alpha_q$.
Due to the zero hub component, none of these states or any linear combination thereof carry current and therefore
\eqref{eq:topStates} does not participate in the scattering problem.

At $\mbox{$E=V_0$}$ and finite momentum $k_y$, that is, $\varphi\neq 0$, the wave function $\Psi_\text{II}(x)\text{e}^{\text{i}k_y y}$ in the barrier region is a superposition
of two evanescent modes and the topological states, given by
%
%
%%%%%%%%%%%%%%%%%%
\begin{eqnarray}
\label{eq:topStatesB}
    \Psi_\text{II}(x)&=&
    \left(\begin{array}{c}
        a\,\text{e}^{k_y x}\\
        0\\
        b\,\text{e}^{-k_y x}
    \end{array}\right)
+\sum_{q}\alpha_{q}
    \left(\begin{array}{c}
        \text{e}^{-\text{i}\theta_q}\\
        0\\
        -\text{e}^{\text{i}\theta_q}
    \end{array}\right)
    \text{e}^{{\rm i}q x},
\end{eqnarray}
%%%%%%%%%%%%
%
%
where $\theta_q=\arctan(k_y/q)$.
As argued above,
the vanishing hub component directly implies that this state does not carry
current, hence we have perfect reflection ($r=-1$) and zero transmission $t=0$.
Note that continuity of $(\psi_{\text A}+\psi_{\text B})$ at $x=0$ and $x=d$ only makes it possible to determine two of the (in principle) infinite amount
of parameters $a,b,\{\alpha_q\}$.
We stress the particularity that at $E=V_0$, the transmission $T(\varphi)$ only is non-zero (and equal to 1) at the singular angle $\phi=0$.
At $E=V_0$ there is no tunneling contribution to the current at finite incident angles, which is in striking contrast to the case of
graphene.\cite{neto:2009}

Finally, we confirm to find the same transmission probability (\ref{vtransmission}) when considering a potential step along an arbitrary direction, perpendicular to some vector $\bm{n}$, as expected from the rotational invariance of the low-energy Hamiltonian (\ref{model}).

\section{Electrostatic barrier in the tight-binding model}

\label{sec:num:sol}

So far we have assumed that the electrostatic potential does not
induce scattering between the two valleys, that is, between the
two inequivalent $K$ and $K'$ points of the first Brillouin zone. In
order to examine the validity of this approximation and to study the
effect of a smooth variation of the potential on the transmission, we
also compute numerically the transmission probability through an
electrostatic barrier in the tight-binding model.

In the following we consider the effect of a potential $V(\bm{n}\cdot
\bm{r})$ on the tight-binding lattice, where $\bm{n}$ is an unit
vector in the direction where the potential changes. Although such a
potential is in principle translationally invariant along the
direction perpendicular to $\bm{n}$, the discreteness of the lattice
reduces the translational invariance to a discrete lattice symmetry.
In particular, if $\bm{n}$ is commensurate with the discrete lattice,
the combined system of potential and lattice is translationally
invariant under shifts of $\bm{t}$, where $\bm{t}$ is lattice vector
perpendicular to $\bm{n}$.  The wave functions are then of the Bloch
form $\psi(\bm{r}) = \text{e}^{\text{i} k_{||} \bm{t}\cdot\bm{r}/
\left|\bm{t}\right |} u(\bm{r})$ with $u(\bm{r}+\bm{t})=u(\bm{r})$
and $k_{||} \in [-\frac{\pi}{\left |\bm{t}\right |}, \frac{\pi}{\left|
\bm{t}\right |})$. The original problem of an infinitely extended
system can thus be reduced to a finite unit cell of width $\left|
\bm{t}\right |$ with periodic boundary conditions in the direction
of $\bm{t}$. The problem must then be solved for each $k_{||}$
individually, with hoppings across the unit cell boundaries being
multiplied by a phase factor of $\text{e}^{\pm \text{i} k_{||}
\left|\bm{t}\right |}$.\cite{noteperiodic} As long as the Fermi
momentum $k_\text{F}$ in the leads (measured with respect to the $K$
and $K'$-points) is smaller than $\frac{\pi}{\left |\bm{t}
\right|}$, the scattering states have conserved parallel momentum in the
leads that allows us to compute the transmission probability $T$ as
a function of the incident angle $\varphi$ as in the continuum case.
The transmission itself is computed using the numerical method of
Ref.~[\onlinecite{wimmer:2009}].

%
%
%%%%%%%%%%%%%%%%

\begin{figure}
   \includegraphics[width=\linewidth]{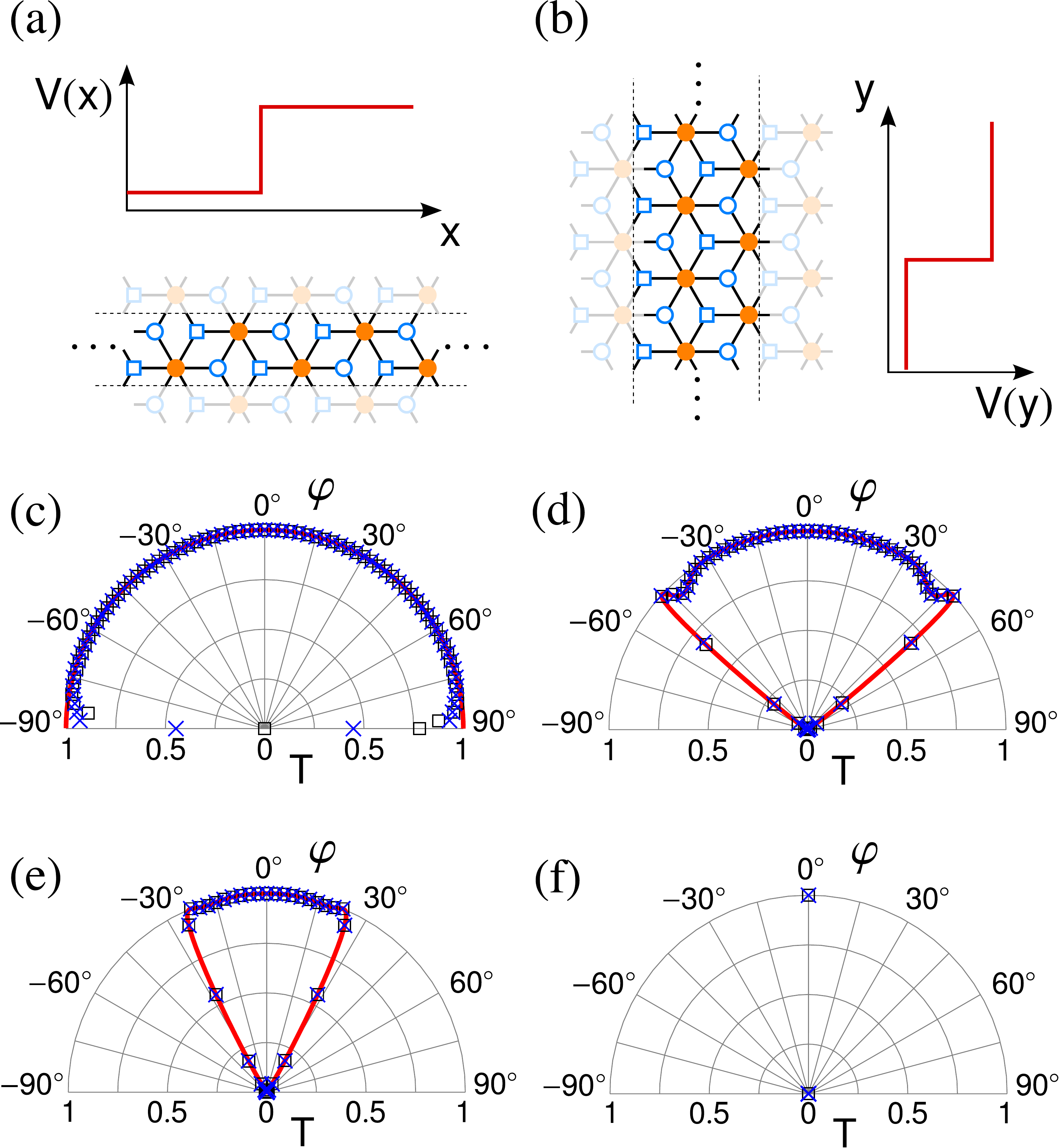}
   \caption{\label{fig:transportNum} Comparison of numerical transport
     simulations and the analytical result for the continuum
     limit. Panels (a) and (b) show the geometries used in the
     numerical simulations, with the potential step along either (a)
     a zigzag direction or (b) an armchair direction.  Panels
     (c)--(f) show a comparison of the transmission $T$ as a function
     of the incident angle $\varphi$ as computed numerically from the
     tight-binding model (black squares, armchair; blue crosses,
     zigzag) and from the continuum result \eqref{vtransmission} [red
     line, left out for clarity in (f), where there is only perfect
     forward scattering at $\varphi=0$]. The parameters used are
     $E=0.01 t$, $d=1200\sqrt{3}a_0$, $d_\text{s}=\sqrt{3}a_0$, and
     (c) $V_0=0.02t$, (d) $V_0=0.018t$, (e) $V_0=0.015 t$, and (f)
     $V_0=0.01 t$.}
\end{figure}

%%%%%%%%%%%%%%%%

%

%

We consider two orientations of the potential step: A step along the
$y$-direction [$\bm{t}=\mathbf{v}_1-\mathbf{v}_2$ with
$\mathbf{v}_{1,2}$ the lattice vectors from
Fig.~\ref{fig:Tau3lattice}(a)] and along the $x$-direction
($\bm{t}=2\mathbf{v}_1+\mathbf{v}_2$) which we --- borrowing from
the HCL nomenclature --- call ``zigzag" and ``armchair" directions,
respectively. Those two geometries, together with the effective unit
cells, are depicted in Figs.~\ref{fig:transportNum}(a) and
\ref{fig:transportNum}(b). While we only show results for two particular
directions, we have checked that orientations in between give
equivalent results.

The two orientations zigzag and armchair are fundamentally different
on the lattice level: A step along a zigzag direction cannot scatter
the valleys, as the $K$ and $K'$-points are at different transverse
(Bloch) momenta $k_{||}$. In contrast, for a step along an armchair
direction both $K$ and $K'$-point are projected to $k_{||}=0$, hence
intervalley scattering is always possible. Note that while an abrupt
step in the zigzag direction cannot scatter the valleys, there still
may be corrections to the low-energy Hamiltonian \eqref{model} in the
form of mass-like terms, as an abrupt step makes the lattice sites
inside a unit cell inequivalent (for an extensive discussion of
equivalent effects in the HCL, see Ref.~[\onlinecite{Tang:2008}]).

In order to compare our result \eqref{vtransmission} from the
low-energy Hamiltonian to the lattice results, we restrict ourselves
to small energies $E \ll t$ and barrier heights $V_0 \ll t$, where $t$
is the hopping matrix element of the tight-binding lattice. In
addition, instead of abrupt jumps we ramp the potential linearly over
a distance $d_\text{s}$ with $k_\text{F} d_\text{s} \ll 1$. A
comparison between our analytical and numerical results is shown in
Figs.~\ref{fig:transportNum}(c)--\ref{fig:transportNum}(f). We find perfect
agreement between the low-energy, continuum and lattice results. As
predicted from the rotational invariance of the low-energy
Hamiltonian, the transmission probability is independent from the
direction of the potential step.  In particular, we find a virtually
angle-independent transmission for $E =0.5 V_0$; small deviations are
only visible for angles close to grazing incidents, where lattice
effects become important. We also find strictly forward transmission
in the case of $E=V_0$ as predicted by the low-energy theory.

Barrier transmission on the \ttt-lattice differs not only at low
energies from graphene, as seen from the different functional
forms of the transmission probabilities $T(\varphi)$
[Eqs.~(\ref{vtransmission}) and (\ref{eq:T:HCL})], but pronounced
dissimilarities arise also at higher energies as investigated
exemplarily now.

\begin{figure}
   \includegraphics[width=\linewidth]{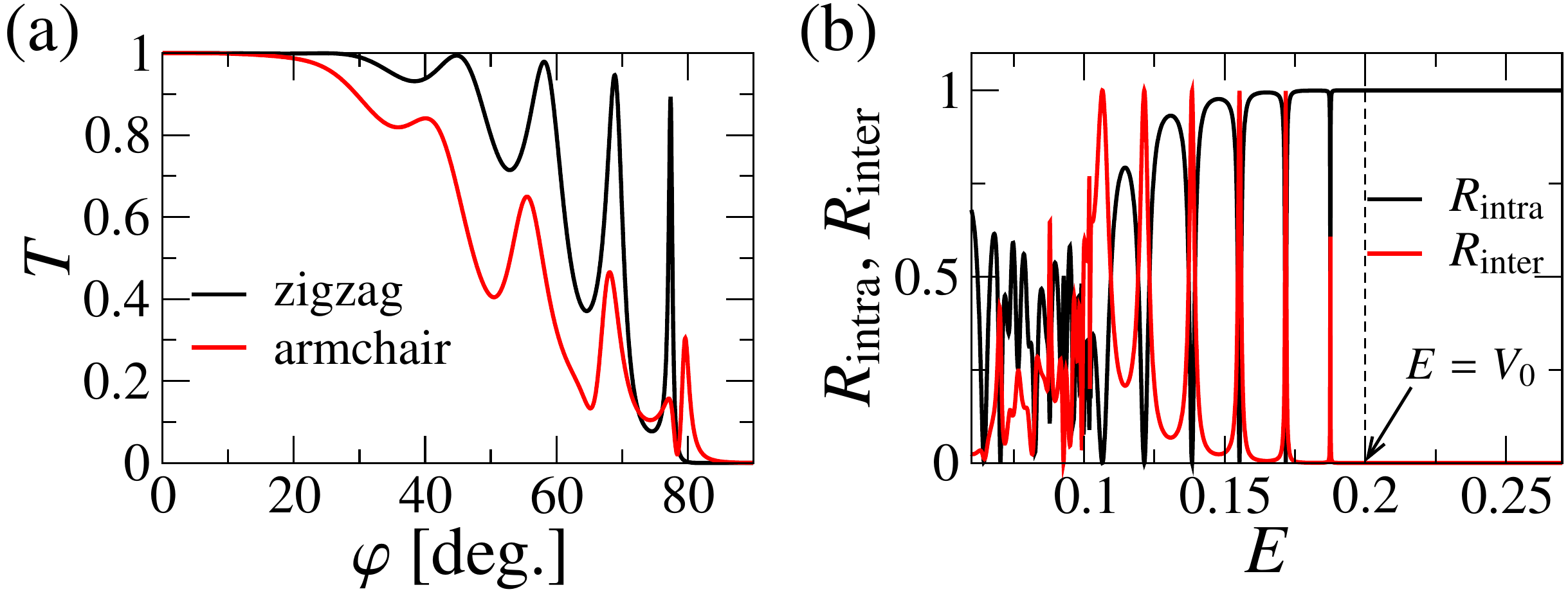}
   \caption{\label{fig:smooth} (a) Transmission as a function of
     incident angle $\varphi$ through a smoothed potential step along
     a zigzag (black line) and armchair direction (red
     line). Parameters are $E=0.1 t$, $V_0=0.2 t$, $d=200
     \sqrt{3}a_0$, and $d_\text{s}=5.5 \sqrt{3}a_0$. (b) Probability for a valley
     flip in the scattering process (given by the total probability of transmission or
     reflection into the opposite valley, $T_\text{inter}+R_\text{inter}$) for a smoothed potential
     step along the armchair direction at the incident angle
     $\varphi=70^\circ$ as a function of $E$. The remaining parameters
     are as in (a).}
\end{figure}

While we found that the transmission probability was independent of
the lattice direction in the case of a small potential step at low
energies, we find a quite pronounced angle dependence at higher
energies and larger potential steps or larger $d_\text{s}$, as seen in Fig.~\ref{fig:smooth}a
for parameters $E=V_0/2$ where the continuum theory yields unit
transmission probability.  Furthermore, for a potential step in the
armchair direction the transmission is considerably smaller than for a
step along the zigzag direction.

We can ascribe the difference to intervalley-scattering:
A step in the zigzag direction still excludes intervalley
scattering due to translational symmetry but it may take place at a
step in armchair direction.

In Fig.~\ref{fig:smooth}b we investigate intervalley scattering
in more detail for an armchair barrier and show the probability for a
valley flip in the scattering process, that is, the probability of transmission or
reflection into the other valley. For energies $E$ in the range $0<E<V_0$
pronounced resonances are seen where intervalley scattering is
strongly enhanced.  This effect even increases with increasing
$d_\text{s}$ (not shown), in stark contrast to the case of a HCL
lattice, where smoother potential barriers reduce intervalley
scattering. We attribute this effect as being mediated by
crossing of the Fermi energy with the flat band connecting both valleys when
$V(\bm{r})=E$ for some $\bm{r}$ and this mediation taking place
more efficiently at smoother barrier slopes (similar as in a
Landau-Zener transition). For energies $E>V_0$, when this
crossing does not take place anymore, intervalley scattering is
suppressed, accordingly.

\section{Uniform magnetic field\label{sec:LLs}}
Now we turn to the effects of a perpendicular magnetic field $B_z(x,y)$ on ${\cal T}_3$.
In the homogeneous case, $B_z(x,y)=B_0$, Landau levels (LLs) appear,~\cite{bercioux:2009} which resemble the relativistic Landau splitting of
graphene in their proportionality to $\omega_{\rm c}\propto\sqrt{B_0}$,
where $\omega_{\rm c}=\sqrt{2}v_F/\ell_B$ is the cyclotron frequency and
$\ell_B=\sqrt{c/(eB_0)}$ the magnetic length.
However, the non-zero Landau energies for the \ttt\ lattice are
%
%
%%%%%%%%%%%%%
\begin{equation}\label{en:tau3}
    E_n^{\ttm}=\pm\omega_\text{c}\sqrt{n-\frac{1}{2}}~~~~ \text{for} ~~~ n\ge 1
\end{equation}
%%%%%%%%%%%%%%
%
%
and differ crucially from their values of graphene given by
$\:E_n^\text{HCL}=\pm\omega_\text{c}\sqrt{n}\:$. We mention
that this latter spectrum is basically known from massless
Dirac-Fermions \cite{relativisticLL} moving at the velocity of light;
it can be attributed to a Zeeman-contribution $\pm ecB_0$
for the two spin helicities,
according to which the relativistic LLs at positive energies
take the values $\:E_{\{{n\atop
n-1}\}}=\omega_\text{c}\sqrt{n-\frac{1}{2}\pm\frac{1}{2}}\:$.
As a result, each LL exhibits a doubled Zeeman-degeneracy with
the exception of the zero-energy level $E_0$ which remains
non-degenerate.~\cite{gusynin:2005} The similar mathematical
structure of the Dirac-Weyl equation for graphene causes
$E_n=E_n^\text{HCL}$ together with only one pseudospin
projection in the zeroth LL which is responsible for the
anomalous Hall conductivity~\cite{gusynin:2005} of graphene, as
observed in experiment.~\cite{QHE:exp} On the other hand, the
spectrum~\eqref{en:tau3} arises due to the \emph{absence} of a pseudo-Zeeman
contribution in the ${\cal T}_3$ case.
It is a signature of the pseudospin 1 physics~\cite{Green:2010,Lan:2011,Goldman:2011,Dora:2011}
and has not yet appeared in other physical
systems.

In Landau gauge, $\bm{A}=(-B_0y,0,0)^{\text T}$, the LLs of ${\cal T}_3$ are given as
%
%
%%%%%%%%%%%%%%%
\begin{equation}
\label{tau3ll}
    \psi_n=\frac{1}{\sqrt{2(2n-1)}}
    \begin{pmatrix}
        \sqrt{n-1}\;\phi_{n-2}\\
        \text{sign}(E_n)\sqrt{2n-1}\;\phi_{n-1}\\
        \sqrt{n}\;\phi_{n}
    \end{pmatrix}\;,\quad n\ge 1
\end{equation}
%%%%%%%%%%%%%%%
%
%
where the $\phi_n$ are eigenstates of the harmonic oscillator
at frequency $\omega_{\rm c}$ and we define $\phi_{-1}\equiv 0$.
While one half of the probability density resides on the hub sites, as expected from
the lattice structure, the remaining half is shared unequally
among the lattice sites A and B according to (\ref{tau3ll}) for
all $n\ge 1$ at finite magnetic field in the vicinity of the
given $K$-point. Again, this contrasts to graphene where A and
B sublattices carry equal probability density for the non-zero LLs.

At zero energy two contributions add to the density of
states: First, the zero-energy LL of degeneracy
equal to the number of flux quanta penetrating the lattice which
equals the degeneracy of all other LLs. The corresponding wave
function,
%
%
%%%%%%%%%%%%%%%
\begin{equation*}
    \psi_0=\left(0, 0, \phi_0\right)^\text{T}
\end{equation*}
%%%%%%%%%%%%%%%%%
%
%
resembles the wave function of graphene's zeroth LL.
On the other hand, there is the ``non-magnetic'' pseudospin
$S_{\bm{p}}=0$ topological level whose energy remains unaffected
when a magnetic field is switched on.
Its degeneracy equals the
number of elementary cells in the lattice and the corresponding wave functions read
%
%
%%%%%%%%%%%%%%%%%%%%%
\begin{equation*}
    \psi_n^\text{top}=\frac{1}{\sqrt{2n+1}}
    \begin{pmatrix}
        \sqrt{n+1}\:\phi_{n-1}\\
        0\\
        -\sqrt{n}\:\phi_{n+1}
    \end{pmatrix}\;,\quad n\ge 1\;.
\end{equation*}
%%%%%%%%%%%%%%%%
%
%
%
%

\section{Magnetic barrier}\label{Sec:MagneticBarrier}

From graphene is known~\cite{demartino:2007} that
magnetic barriers can confine Dirac-Weyl fermions with zero
transparency, contrary to electrostatic barriers.
We consider a barrier parallel to the $y$-axis of thickness $d$ given by a space-dependent magnetic field perpendicular to the lattice plane,
%
%
%%%%%%%%%%%%%%%%
\begin{equation}\label{magbarrier}
    B_z(x,y)=B_0\Theta(d/2-|x|)\;,
\end{equation}
%%%%%%%%%%%%%%%
%
%
described by a corresponding vector potential ${\bm A}(x)=\{0,A_y(x),0\}$,
%
%
%%%%%%%%%%%%
\begin{equation*}
A_y=B_0\left\{\begin{array}{ccc}
-d/2&&x<-d/2\\
x&\;{\rm for}\;&|x|\le d/2\\
d/2&&x>d/2
\end{array}\right.\;.
\end{equation*}
%%%%%%%%%%%%%%
%
%
Stationary states  of the Hamiltonian (\ref{model}) of the form \eqref{general:wf} satisfy the set of coupled
differential equations
%
%
%%%%%%%%%%%%%%%%%%%%
\begin{subequations}
\label{DiracEq:Finite:B}
\begin{align}
    \sqrt{2}{\cal E}\psi_{\text A}&=
    -\text{i}\partial_x\psi_{\text H}-\text{i}\left(k_y+\frac{e\,\ell_B^2}{c}A_y(x)\right)\psi_{\text H},
    \\
    \sqrt{2}{\cal E}\psi_{\text H}&=
    -\text{i}\partial_x\psi_{\text A}+\text{i}\left(k_y+\frac{e\,\ell_B^2}{c}A_y(x)\right)\psi_{\text A}\nonumber\\
    &\quad-\text{i}\partial_x\psi_{\text B}+\text{i}\left(k_y+\frac{e\,\ell_B^2}{c}A_y(x)\right)\psi_{\text B},
    \\
    \sqrt{2}{\cal E}\psi_{\text B}
    &=
    -\text{i}\partial_x\psi_{\text H}+\text{i}\left(k_y+\frac{e\,\ell_B^2}{c}A_y(x)\right)\psi_{\text H},
\end{align}
\end{subequations}
%%%%%%%%%%%%%%%%%%%
%
%
where $k_y$ is the transverse momentum. Here and in the following we have rescaled all quantities by using $\ell_B=\sqrt{c/(eB_0)}$ as the unit of length and
$v_F/\ell_B=\omega_\text{c}/\sqrt{2}$ as the unit of
energy, ${\cal E}=E/(v_F/\ell_B)$.
The wave function left of the barrier is of the form (\ref{psi1}) with $k_y={\cal E} \sin(\varphi)+d/2$, where
$\varphi$ is again the angle of incidence.
Conservation of momentum parallel to the barrier results in the condition
%
%
%%%%%%%%%%%%%%%%
\begin{equation}
\label{eq:kyconservation:B}
    \sin(\varphi)+d/{\cal E}=\sin(\alpha)
\end{equation}
%%%%%%%%%%%%%%%
%
%
so that the wave function on the right side of the barrier reads
%
%
%%%%%%%%%%%%%%%%%%
\begin{eqnarray}
    \psi_{\text{III}}(x)&=&t\
    %\sqrt{\frac{k_x}{q_x}}\;
        \begin{pmatrix}
        \text{e}^{-{\rm i}\alpha}\\
        \sqrt{2}s\\
        \text{e}^{{\rm i}\alpha}
    \end{pmatrix}\text{e}^{\text{i} q_x x}
\end{eqnarray}
%%%%%%%%%%%%%%%%%
%
%
with emergence angle $\alpha$ and final momentum in the $x$-direction $q_x=\sin(\alpha){\cal E}$.
As in the case of graphene,\cite{demartino:2007} the condition of momentum conservation in the $y$-direction
(\ref{eq:kyconservation:B}) cannot be fulfilled for sufficiently thick barriers, $d>2\mathcal{E}$.
Finally, the wave function in the barrier region is given by
%
%
%%%%%%%%%%%%%%%%
\begin{eqnarray}
    \psi_{\text{II}}(x)&=&
    a\;\begin{pmatrix}
        \xi \mathcal{D}_{\eta+1}\left(\xi\right)
        -\mathcal{D}_{\eta}
        \left(\xi\right)
    \\
        \text{i} \mathcal{E}
   \mathcal{D}_{\eta+1}\left(\xi\right)
   \\
   -\mathcal{D}_{\eta}\left(\xi\right)
   \end{pmatrix}
   \nonumber\\&&\quad+
   b\;\begin{pmatrix}
        -\mathcal{D}_{1-\eta }(\text{i}\xi)\\
        \mathcal{E} \mathcal{D}_{-\eta}(\text{i}\xi)\\
        \text{i} \xi \mathcal{D}_{-\eta }(\text{i}\xi)-\mathcal{D}_{1-\eta}(\text{i}\xi)
   \end{pmatrix},
\end{eqnarray}
where the ${\cal D}_\eta$ are parabolic cylinder functions taken at argument
$\xi=\sqrt{2}(k_y+x)$ and $\eta=( \mathcal{E}^2+1)/2$.

%
%
%%%%%%%%%%%%%%
\begin{figure}
	\centering
	\includegraphics[width=0.45\textwidth]{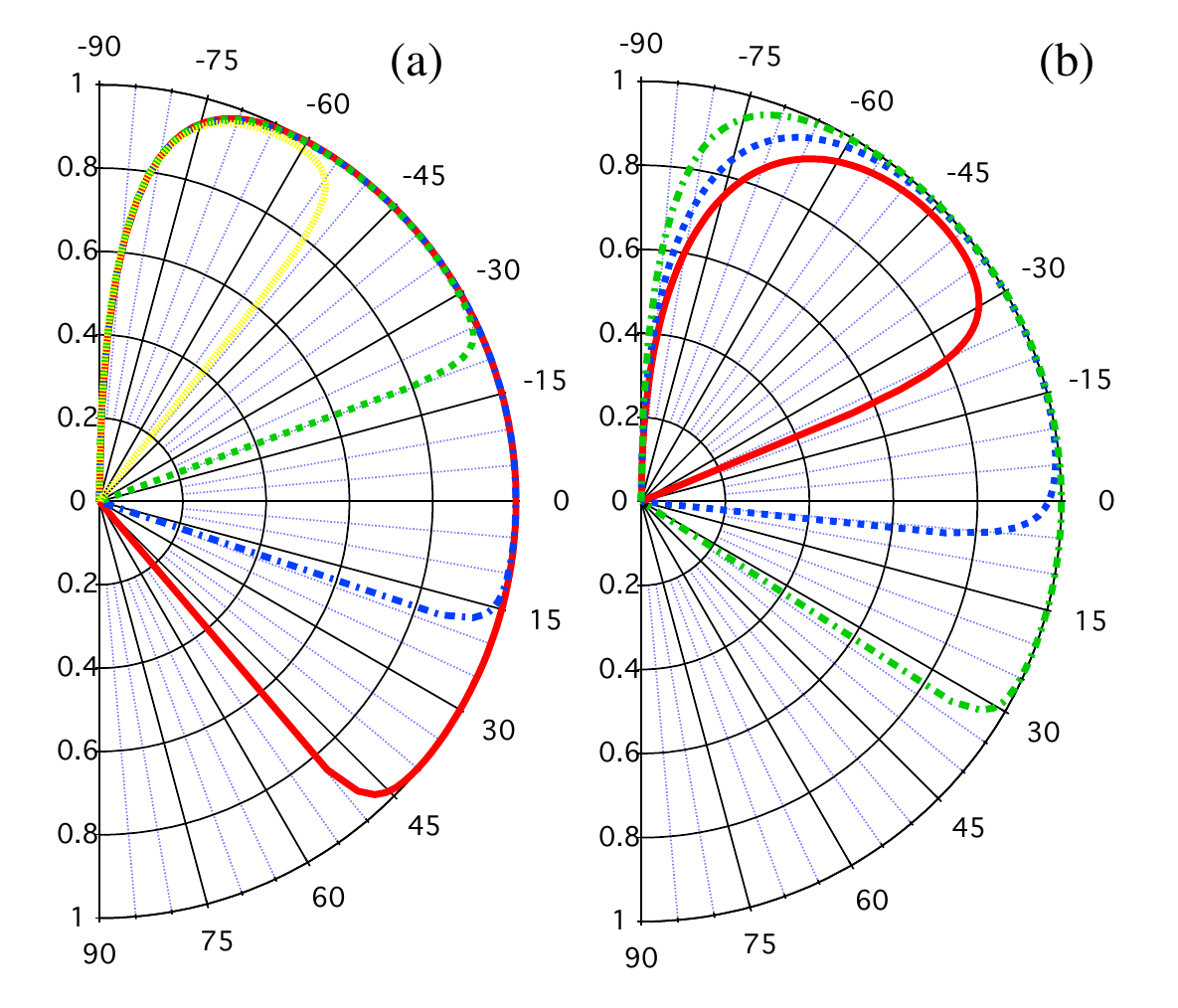}
	\caption{\label{fig:ResultB:PolarPlot} (Color online). The transmission probability $T$ as a function of the injection angle for a magnetic barrier of width $d$ and at the energy $\mathcal{E}$ for the \ttt lattice. In panel (a) the parameters are $\mathcal{E}=4.5$, $d=1$ (red solid line), $d=3$ (blue dot-dashed line), $d=6$ (green dashed line), and $d=8$ (yellow dotted line). In panel (b) the parameters are $d=2.2$, $\mathcal{E}=1.6$ (red-solid line), $\mathcal{E}=2.5$ (blue dotted line), and $\mathcal{E}=5$ (green dot-dashed line).}
\end{figure}
%%%%%%%%%%%%%%
%
%
%%%%%%%%%%%%%%
\begin{figure}
	\centering
	\includegraphics[width=\columnwidth]{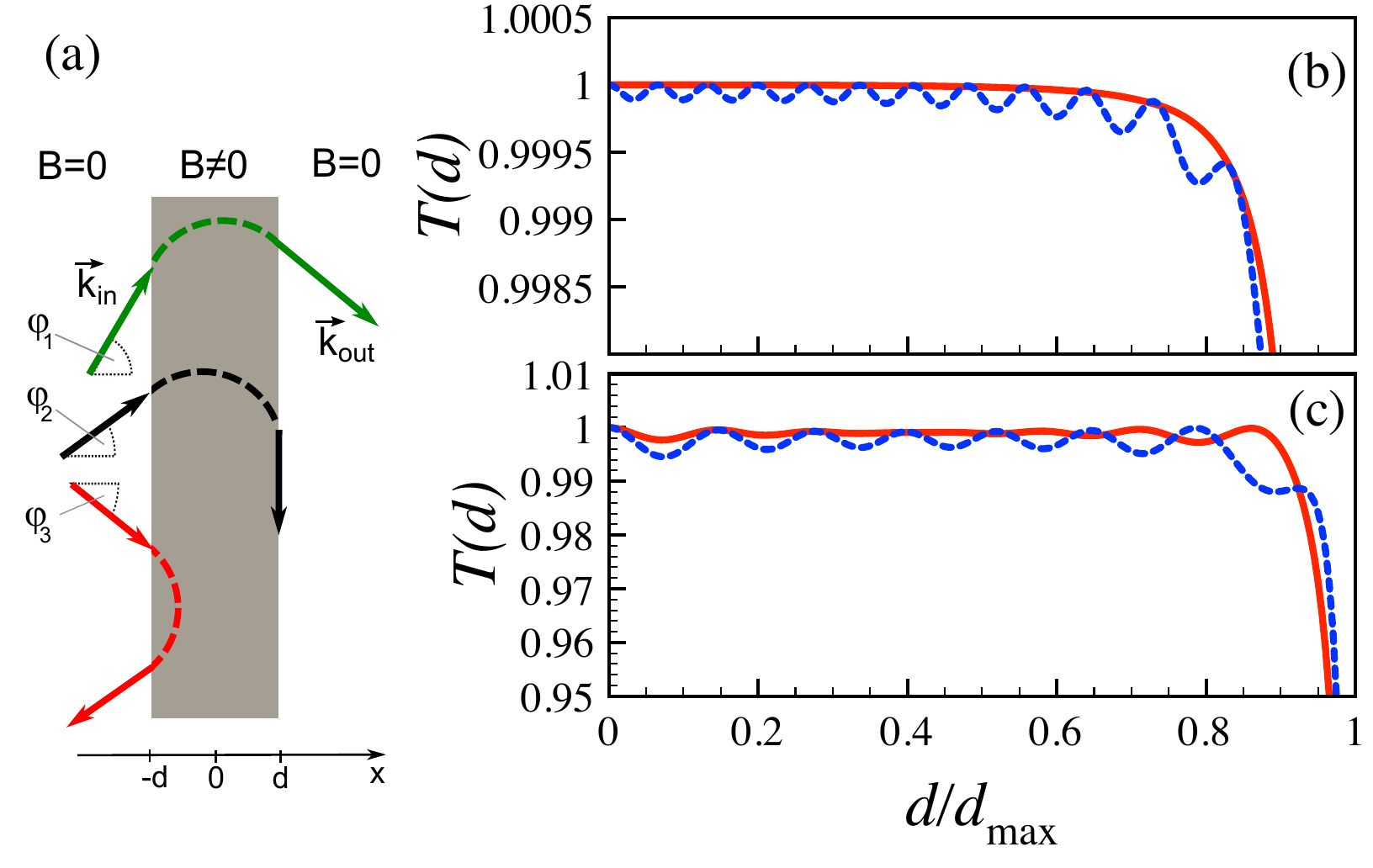}
	\caption{\label{fig:resultsB:Zoom} (Color online). (a) Sketch of possible semiclassical trajectories of particles entering the barrier region at different incident angles, as explained in the text. Panels (b) and (c): transmission probability as a function of the normalized width of the barrier $d/d_\text{max}$ for the \ttt lattice (solid red lines) and HCL (dashed blue lines). Panel (b): $\mathcal{E}=6.9$ and $\varphi=-\pi/4$. Panel (c): $\mathcal{E}=3.9$ and $\varphi=0$.}
\end{figure}
%%%%%%%%%%%%%%
%
The transmission probability $T=|t|^2$ can be calculated
straight-forwardly by applying the boundary conditions
(\ref{matching}). The exact result is lengthy and given in
Appendix~\ref{App:ResultB}; it is plotted in
Fig.~\ref{fig:ResultB:PolarPlot}. The main features of this
result can be understood within a semiclassical picture,
according to which electrons perform cyclotron orbits of radius
$r_0=(E/v_{\rm F})\ell_B^2$, inversely proportional to the strength
of the magnetic field.\cite{CyclotronOrbitHCL} The semiclassical
approximation is justified for $r_0\gg\ell_B$. Figure
\ref{fig:resultsB:Zoom}(a) illustrates three possible cases. (i)
For an incident angle $\varphi_1>\varphi_\text{crit}$ larger
than a critical angle the particle reaches the opposite boundary
and leaves with transmission probability $T=1$. (ii) At the
value of the critical angle, $\varphi_2=\varphi_\text{crit}$,
the particle leaves the barrier region parallel to the barrier.
(iii) For angles $\varphi_3<\varphi_\text{crit}$, the particle
makes the full turn under the barrier and is perfectly
reflected, $T=0$. In this semiclassical picture the critical
angle $\varphi_\text{crit}$ is obtained from
%
%
%%%%%%%%%%%%%%%%%
\begin{equation}
\label{eq:phi:crit}
    d_\text{max}={\cal E}(1+\sin(\varphi))
\end{equation}
%%%%%%%%%%%%%%%%%%
%
%
which relates the maximal barrier width $d_\text{max}$ that can be overcome at incident angle $\varphi$ and energy $\cal E$.
In fact, the curves obtained for the HCL, taking the results of Ref.~[\onlinecite{demartino:2007}], are almost identical when plotted on the scale of Fig.~\ref{fig:ResultB:PolarPlot},
confirming the semiclassical point of view also for graphene.

In Figs.~\ref{fig:resultsB:Zoom}(b) and \ref{fig:resultsB:Zoom}(c) we show the transmission
probabilities for both lattices as a function of the
normalized magnetic barrier width $d/d_\text{max}$ at
different angles and energies where the transmission for
$d<d_\text{max}$ is very close to unity. However, the \ttt
lattice reveals as more transparent than the HCL, so that the
latter is seen to develop quantum interference oscillations as a
function of the barrier width.

\section{Summary}

We have analyzed transmission properties of barriers on the \ttt\ lattice
in the low energy approximation using the Dirac-like Hamiltonian (\ref{model}).
First we derived and discussed the boundary conditions for the wave function at a given interface. Wave function components
of rim sites can be discontinuous, whereas the hub component is continuous; this can be well understood in terms of the probability currents perpendicular and parallel to the barrier.
An electrostatic barrier on \ttt\ has enhanced transparency compared to the HCL, which is a direct consequence
of the specific boundary conditions of the $S=1$ Hamiltonian.
When the energy of the incoming wave equals half the barrier hight, we even obtain perfect transmission independently of the incident angle.
This ``super Klein-tunneling'' makes it possible to use a $np$-junction in order to design a perfect focusing
lens without loss; this must be contrasted to the 50$\%$ of loss of HCL in the same configuration.
Therefore, we believe that \ttt\ can be interesting from the point of view of electron focusing and photonic crystals.\cite{Joannopoulos:2008}

Furthermore, we have confirmed numerically the analytical predictions about the super Klein-tunneling via a lattice Green's function method. In addition, we have investigated numerically the influence of the intervalley scattering in the \ttt\ lattice compared to the HCL and identified the crucial role of the flat band.

For a uniform magnetic field we have discussed the LLs;
their energies differ from non-relativistic and from relativistic
systems known so far. \ttt\ exhibits a high density of states at
zero energy due to contributions from a zeroth LL and
a topological level.

Finally, we have investigated the transmission through a magnetic barrier in \ttt\ which is found to be similar compared to the case of the HCL.
Both cases can be understood in a simple semiclassical picture
from circular cyclotron orbits under the barrier. Magnetic barriers of sufficient width can be used to confine particles in the \ttt\-lattice.

\begin{acknowledgements}
We acknowledge A.~De Martino, H.~Grabert, V.~Kr\"uckl, and P.~Recher for useful discussions. The work of DB is supported by the Excellence Initiative of the German Federal and State Governments.
\end{acknowledgements}
\appendix

\section{On the boundary conditions\label{App:one}}

We can obtain the boundary conditions from the low energy wave equation
\eqref{model} together with the conservation of the probability current. The former states that $-\ii\partial_t \Psi = \mathcal{H} \Psi$, the latter that $\partial_t |\psi|^2 = -\bm{\nabla}\cdot\bm{j}$.

For the case of graphene and with the two component spinor $\Psi=(\psi_\text{A},\psi_\text{B})^\text{T}$ we get the following condition:
%
%
%%%%%%%%%%%
\begin{subequations}\label{eq:graphene}
\begin{align}
\partial_t|\Psi|^2 & = (\partial_t \Psi) \Psi^*+(\partial_t \Psi^*) \Psi \\
& = -2v_\text{F} \partial_x {\rm Re}[\psi_\text{A}^*\psi_\text{B}]-2v_\text{F} \partial_y {\rm Im}[\psi_\text{A}^*\psi_\text{B}]\\
& = -(\partial_x j_x + \partial_y j_y)\,.
\end{align}
\end{subequations}
%%%%%%%%%%%
%
%
Therefore, we can identify the two components of the probability current
%
%
%
%%%%%%%%%%%%%%%
\begin{equation}
\bm{j}=2v_\text{F}({\rm Re}[\psi_\text{A}^*\psi_\text{B}],{\rm Im}[\psi_\text{A}^*\psi_\text{B}])^\text{T}.
\end{equation}
%%%%%%%%%%%%%%
%
%
At an interface we require the continuity of the probability current $j_x$ perpendicular to the boundary. This implies for the HCL the continuity of the two components of the wave function $\Psi$ at the interface.

On the other hand, for the case of the \ttt\ lattice and the three-component spinor $\Psi=(\psi_\text{A},\psi_\text{H},\psi_\text{B})^\text{T}$, we get
%
%
%%%%%%%%%%%
\begin{subequations}
\begin{align}
\partial_t|\Psi|^2 & = (\partial_t \Psi) \Psi^*+(\partial_t \Psi^*) \Psi \\
& = -\sqrt{2}v_\text{F} \partial_x {\rm Re}[\psi_\text{H}^*(\psi_\text{A}+\psi_\text{B})] \nonumber \\
& \hspace{1cm} +\sqrt{2} v_\text{F} \partial_y {\rm Im}[\psi_\text{H}^*(\psi_\text{A}-\psi_\text{B})]\\
& = -(\partial_x j_x + \partial_y j_y)\,.
\end{align}
\end{subequations}
%%%%%%%%%%%
%
%
In this case the probability current is defined by
%
%
%%%%%%%%%%%%%%%%%%%%%
\begin{equation}
\bm{j}=\sqrt{2}v_\text{F}({\rm Re}[\psi_\text{H}^*(\psi_\text{A}+\psi_\text{B})],{\rm Im}[\psi_\text{H}^*(\psi_\text{A}-\psi_\text{B})])^\text{T}.
\end{equation}
%%%%%%%%%%%%%
%
%
The conservation of the $x$ component of the current corresponds indeed to Eqs.~\eqref{matching}.

\section{Transmission probability through a magnetic barrier}
\label{App:ResultB}
The transmission probability through the magnetic barrier discussed in Sec.\ \ref{Sec:MagneticBarrier}
is obtained by imposing the matching conditions \eqref{matching} at $x=\pm d/2$ that after basic algebraic manipulation read
%
%
%%%%%%%%%%%%%%%%
\begin{eqnarray}
    a \mathcal{D}_{\eta-1}'\left(\xi_-\right)+b \mathcal{D}_{-\eta }'\left(\text{i}\xi_-\right)&=&\cos(\varphi )\left(\text{e}^{-\text{i} \frac{d}{2} k_x}+r \text{e}^{\text{i} \frac{d}{2} k_x}\right) \text{e}^{\text{i} k_y y},
\nonumber\\
    a \mathcal{D}_{\eta-1}'\left(\xi_+\right)+b \mathcal{D}_{-\eta }'\left(\text{i} \xi_+\right)&=&t \cos(\alpha)\text{e}^{\text{i} (\frac{d}{2} q_x+q_y y)},
\nonumber\\
    \text{i} a \mathcal{D}_{\eta-1}\left(\xi_-\right)+b \mathcal{D}_{-\eta }\left(\text{i} \xi_-\right)
    &=&\frac{\sqrt{2}}{|\mathcal{E}|}\left(\text{e}^{-\text{i} \frac{d}{2} k_x}-r \text{e}^{\text{i} \frac{d}{2} k_x}\right) \text{e}^{\text{i} k_y y},
\nonumber\\
   \text{i} a \mathcal{D}_{\eta-1}\left(\xi_+\right)+b
   \mathcal{D}_{-\eta }\left(\text{i} \xi_+\right)
    &=&\frac{\sqrt{2}}{|\mathcal{E}|}t \text{e}^{\text{i} (\frac{d}{2} q_x+q_y y)},
\nonumber
\end{eqnarray}
%%%%%%%%%%%%%%%%
%
%
with $\xi_\pm=\sqrt{2}(k_y\pm d/2)$. The prime denotes the derivative with respect to the coordinate. The expression for the resulting transmission is lengthy. It can be written in a more compact form when introducing the function
%
%
%%%%%%%%%%%%%%%%
\begin{eqnarray}
    G_\eta(\xi_1,{\xi}_2):=\mathcal{D}_{-\eta}(\text{i}\xi_1)\mathcal{D}_{\eta-1}({\xi_2})-
    \mathcal{D}_{\eta-1}(\xi_1)\mathcal{D}_{-\eta}(\text{i}{\xi_2})\nonumber\\
\end{eqnarray}
%%%%%%%%%%%%%%%%%
%
%
and its derivatives $\partial_{1/2}G_\eta$ with respect to the first/second argument. Then the final result reads
%
%
%%%%%%%%%%%%%%%%
\begin{widetext}
\begin{eqnarray}
    T&=&\left|\frac{2\sqrt{2}{|\mathcal{E}|}\cos(\varphi)\partial_1G_\eta(\xi_+,\xi_+)}{
    \mathcal{E}^2G_\eta(\xi_+,\xi_-)
    +\text{i}\sqrt{2}|\mathcal{E}|\left[\cos(\alpha)\partial_2G_\eta(\xi_+,\xi_-)
    -\cos(\varphi)\partial_1G_\eta(\xi_+,\xi_-)\right]
    +2\partial_1\partial_2G_\eta(\xi_+,\xi_-)
    }\right|^2.
\end{eqnarray}
\end{widetext}
%%%%%%%%%%%%%%%
%
%


\begin{thebibliography}{99}
\bibitem{noteperiodic} Formally, the problem is then equivalent to a nanotube geometry with a flux
of $k_{||} \left |\bm{t}\right | \Phi_0$ inside the tube.
\end{thebibliography}
\end{document}